\documentstyle[prl,aps,epsf]{revtex}
\begin{document}
\draft
\title{The $1/N$-expansion, quantum-classical correspondence and
           nonclassical states generation in dissipative
           higher-order anharmonic oscillators}
\author{Kirill N. Alekseev$^{a,b}$\cite{email1},
and Jan Pe\v{r}ina$^{a}$\cite{email2}}
\address{
$^a$Department of Optics and Joint Laboratory of Optics of\\
 Palack\'{y} University and Academy of Sciences of Czech Republic, \\
17. listopadu 50, 772 07 Olomouc, Czech Republic\\
$^b$Theory of Nonlinear Processes Laboratory,
Kirensky Institute of Physics,\\
Russian Academy of Sciences, Krasnoyarsk 660036, Russia}
\maketitle
\begin{abstract}
We develop a method for the determination of thecdynamics of dissipative
quantum systems in the limit of large number of quanta $N$, based on the
$1/N$-expansion of Heidmann {\it et al.} [ Opt. Commun. {\bf 54}, 189 (1985) ]
and the quantum-classical correspondence. Using this method, we find
analytically the dynamics of nonclassical states generation in the higher-order
anharmonic dissipative oscillators for an arbitrary temperature of a reservoir.
We show that the quantum correction to the classical motion increases with
time quadratically up to some maximal value, which is dependent on the degree
of nonlinearity and a damping constant, and then it decreases. Similarities
and differences with the corresponding behavior of the quantum corrections to
the classical motion in the Hamiltonian chaotic systems are discussed.
We also compare our results obtained for some limiting cases with the results
obtained by using other semiclassical tools and discuss the conditions for
validity of our approach.
\end{abstract}
\pacs{03.65.Sq, 42.50.Dv, 42.50.Ne}
\section{Introduction}
\label{sec:introduc}
The quantum anharmonic oscillator with the Hamiltonian in the interaction
picture ($\hbar\equiv 1$)
\begin{equation}
\label{1}
H=\Delta b^{\dag}b + \frac{\lambda_l}{l+1} \left( b^{\dag} b \right)^{l+1},
\quad [ b, b^{\dag} ]=1
\end{equation}
is one of the simplest and the most popular models describing the quantum
statistical properties of light interacting with a nonlinear medium \cite{1,2}.
In Eq. (\ref{1}), the operators $b$ and  $b^{\dag}$ describe a single mode of
quantum field and
the constant $\lambda_l$ is proportional to the $(2 l+1)$-order nonlinear
susceptibility of a nonlinear medium ($l$ is an integer), $\Delta$ is
the detuning of the light frequency from the characteristic frequency of
quantum transition. We adopt the normal ordering of operators.
For the case of a cubic nonlinearity ($l=1$), this model first was  introduced
by Tana\'{s} \cite{3} for the investigation of self-squeezing of light
propagating through a nonlinear Kerr medium without loss.
Because of such a model is exactly integrable, the explicit time dependence of
the quadrature variances necessary for the
determination of squeezing condition has been found for any
moments of time and for
any number of photons \cite{3,4}.  The problem of a
dissipative anharmonic oscillator is much more difficult.
Nevertheless, Milburn and Holms obtained the
exact solution  for the damped Kerr oscillator ($l=1$) interacting with
a reservoir
of zero temperature \cite{5}. This result has been further generalized to
the case of a reservoir of non zero temperature in \cite{6}.
\par
In the conditions of an
experiment, as a rule, a large number of photons are involved in a nonlinear
interaction between light and a nonlinear medium modelled by the anharmonic
oscillator \cite{1,3,4}.
The determination of squeezing conditions from the exact solution in this
limiting case is straightforward for the model of Kerr oscillator without
loss \cite{4}.
In contrast, due to the complex form of the exact solution for the damped Kerr
oscillator,
the determination of photon statistics for the large number of photons in
this model
demands an application of numerical methods or special approximate
analytical methods (for a review see \cite{2}).
Moreover, there are no exact solutions for the model of the quantum
dissipative oscillator with higher-order nonlinearity and a very little
amount of the information on its dynamics is documented in the literature
\cite{2}.
\par
In general, the situation when a large number of photons $N$ are involved
in nonlinear interactions is a quite typical for many problems of
quantum optics \cite{reynaud,fabre}. Heidmann {\it et al.} suggested
\cite{7} to use the method of $1/N$-expansion for the determination of
nonclassical states generation dynamics. They originally
applied the $1/N$-expansion technique to the problem of
squeezing and antibunching of an electromagnetic field interacting
with a collection of the Rydberg atoms inside a high-$Q$ cavity
\cite{heidmann-prl}, where a large number of atoms is of the same order as
number of the photons $N$. The general scheme of the $1/N$-expansion method
states that an exact or an approximate solution of the problem
can be found in the classical limit $N\rightarrow\infty$ and then
the quantum corrections could be added \cite{yaffe}. Because this
method allows to find the motion equations for the mean values
and the lower-order cumulants, it could also be considered as a
variant of the cumulant expansion \cite{7,schack}. Recently we
further developed and applied the $1/N$-expansion technique \cite{7}
to the investigation of an enhanced squeezing at the transition
to quantum chaos\footnote{
Recently a good agreement between the predictions for dynamics of squeezing
obtained using the method  of  $1/N$-expansion and the results
of numerical simulation for the model of kicked quantum rotator with
$N\simeq 10^5$ levels were found in the works \cite{jetp,glasgow}. }
\cite{8,9,9'}. It should be noticed that only nondissipative
quantum systems have been considered in papers \cite{7,8,9,9'}.
\par
In this paper, using the method of $1/N$-expansion, we consider a
dynamics of squeezing and a deviation from
the Poissonian statistics for the damped anharmonic oscillators with arbitrary
degree of nonlinearity $l$ in the limit of a large number of photons
$N\gg 1$. We find the explicit time dependencies for the squeezing
and the Fano factor for an arbitrary degree of nonlinearity and
for an arbitrary temperature of a reservoir. Our consideration is
based on the quantum-classical correspondence and the fact that the
solution of classical equations of motion, obtained within the
zero-order approximation of $1/N$, could be found analytically
for the case of any linear damping. We show that for a weak
damping the degree of squeezing is mainly determined by the
nonlinear polarization of a nonlinear medium, modelled by the
nonlinear oscillator. For the case of no damping, our time
dependencies for squeezing are transformed to the corresponding
formulas of work \cite{4}, which have been found from an exact
solution of the Hamiltonian problem. A finite damping decreases
the degree of squeezing. The consideration of the Fano factor
demonstrates that the quantum statistics is always a
super-Poissonian for dissipative oscillators. Another restrictive
factor having influence on the time dependencies of squeezing and
the Fano factor are the thermal fluctuations of the reservoir.
\par
Note that in spite of the fact that we find our main results for the the model
of the higher-order oscillator, we present our self-consistent system
of motion equations for the first- and the second-order cumulants in
a form which is valid for the description of any single-mode quantum
system in the semiclassical limit. One of the main finding of
this general consideration consists in the influence of a specific
quantum diffusive term on the dynamics of the expectation values and
dispersions. We interpret this diffusion, which is proportional
to the damping constant, as an influence of the zero-point energy of
reservoir on the quantum system. Although the influence of quantum diffusion
around the classical solution is insufficient for the description of
time dependencies of squeezing and mean values for the particular system
under study, and especially in the most interesting case of a short time of
interactions, we think that the account of this quantum diffusion is
important for the correct description of other dissipative quantum systems in
the semiclassical limit.
\par
We compare our basic equations of motion for dissipative systems with the
equations arising within the so-called generalized Gaussian
approximation \cite{1,schack,perina1} and find a one-to-one correspondence
up to terms of $1/N^2$ for several popular models of quantum optics
\cite{perina1,szlachetka1,perina2,szlachetka2}.
\par
We also discuss the conditions for validity of the cumulant expansion
in the form of the $1/N$-expansion for the description of the dissipative
dynamics of nonlinear oscillators. This problem is related to the problem
of finding the time interval for the quantum-classical correspondence,
which attracts large attention nowadays, and especially in connection with
the studies of quantum chaotic systems (see
\cite{berman-book} and the references cited therein).
For the Hamiltonian systems with regular dynamics, the quantum corrections
to the corresponding classical equations grow in the course of time
power-wise \cite{berman-zaslavsky,berman-book,sundaram,8,9,9'}.
As a result, the time interval for the classical description has
a power-wise dependence on the semiclassical parameter $N$
\cite{berman-zaslavsky,berman-book,sundaram}. In contrast, for the
case of nondissipative quantum systems which are chaotic in the classical
limit, the quantum corrections grow exponentially in time due to
underlying local instability in the classical system
\cite{berman-zaslavsky,berman-book,sundaram,8,9,9'}. Therefore, the time
interval for the validity of the $1/N$-expansion method and the classical
description is logarithmic in the semiclassical parameter $N$
\cite{berman-zaslavsky,berman-book,sundaram}.
\par
Our finding for the dissipative nonlinear oscillators is that the quantum
correction first increases, then reaches some maximum and finally decreases
in the course of time. As a consequence, the time interval for validity of
the $1/N$-expansion could be divided into two subintervals with completely
different dependencies on the semiclassical parameter $N$. The first
subinterval scales up power-wise with $N$ and the second one has a
logarithmic dependence on $N$. While a power dependence on $N$ originates
from the same time behavior as in the Hamiltonian systems with regular
dynamics, the $\log N$ scale has a different nature. It appears because
of an exponential damping of the underlying classical dynamics.
\par
Our work is organized as follows. We consider a single mode dissipative
quantum system and present the derivation of motion equations for the
first and second-order cumulants in section \ref{sec:Basic_eq}. In the same
section we also compare our approach based on the $1/N$-expansion method
with other semiclassical methods and find analytically the time dependencies
for the mean value and the second-order cumulants for the model of the
higher-order dissipative oscillator. Using these  results, we focus on
the time dependencies of the squeezing and the Fano factor in section
\ref{sec:Squeez_and_Fano}. The section \ref{sec:validity} is devoted to
the discussion of conditions
for the validity of the $1/N$-expansion in the description of
nonlinear oscillators dynamics. In the concluding section, we briefly
summarize our results and outlook the main directions for the future
developments. Some details related to the solution of motion equations
for the cumulants and the determination of time interval for the validity
of our approach are presented in two Appendices.
\section{Basic equations}
\label{sec:Basic_eq}
\subsection{The $1/N$-expansion method and comparison with other
approaches}
\label{subsec:1/N-expan}
First of all, we need to generalize the approach of \cite{7,9,9'} to the case
of a single-mode quantum system with dissipation. Consider an oscillator
with the Hamiltonian (\ref{1}) which interacts with an infinite
linear reservoir of finite temperature. The Hamiltonians of a reservoir and
its interaction with oscillator are the following
\begin{equation}
\label{2}
H_r=\sum_{j} \psi_j (d_j^{\dag}d_j+1/2), \quad
H_{int}=\sum_{j}\left(\kappa_j d_j b^{\dag} + {\rm H. c.}\right),
\end{equation}
where the Bose operator  $d_j$ ($[d_j, d_k^{\dag}]=\delta_{jk}$) describes
an infinite reservoir with characteristic frequencies $\psi_j$,
and $\kappa_j$ are coupling constants between the reservoir modes and
the oscillator. Introduce new scaled operators
$a=b/N^{1/2}$, $c_j=d_j/N^{1/2}$,
and Hermitian conjugats to them with commutation relations
\begin{equation}
\label{4}
[ a, a^{\dag} ]=1/N, \quad [c_j, c_k^{\dag}]=\delta_{jk}/N.
\end{equation}
In the classical limit $N\rightarrow\infty$, we have commuting classical
$c$-numbers instead of operators. Now the full Hamiltonian
$H=H_0+H_r+H_{int}$ may be rewritten
as $H=N {\cal H}$, where ${\cal H}$ has the same form as in the formulas
(\ref{1}) and (\ref{2}) with an account of the following replacements
\begin{equation}
\label{5}
b\rightarrow a,\quad b^{\dag}\rightarrow a^{\dag},\quad
d_j\rightarrow c_j,\quad d_j^{\dag}\rightarrow c_j^{\dag},\quad
\mbox{and}\quad \lambda_l\rightarrow g_l(N)\equiv \lambda N^l.
\end{equation}
It could be shown that dependent on photon number constant $g_l(N)$
correctly
gives the time scale of energy oscillations for the nonlinear oscillator
(\ref{1}) in the classical limit (for the case of the Kerr nonlinearity,
see, {\it e.g.} \cite{10}).
\par
Within the standard Heisenberg-Langevin approach the equation of motion has
the form (\cite{1}, chap. 7; \cite{mandel_wolf})
\begin{equation}
\label{5'}
i\dot{a}=\left(\Delta-i\frac{\gamma}{2}\right) a + V +L(t),
\end{equation}
where $V=\partial {\cal H}_0/\partial a^{\dag}$,
$\gamma=2\pi|\kappa(\omega)|^2\rho(\omega)$ is a damping constant,
$\rho(\omega)$ being the density function of reservoir oscillators,
which spectrum is considered to be flat. The Langevin force
operator $L(t)=\sum_{j}\kappa_j d_j(0)\exp(-i\psi_j t)$
has properties \cite{1,mandel_wolf},
which in our notations (\ref{5}) may be rewritten as
\begin{equation}
\label{6}
\langle L(t)\rangle_R=\langle L^{\dag} (t)\rangle_R=0,\quad
\langle L^{\dag}a\rangle_R+\langle a^{\dag}  L\rangle_R=
\gamma\frac{\langle n_d\rangle}{N},\quad
\langle L a\rangle_R+\langle a L\rangle_R=0,
\end{equation}
where the average is performed over the reservoir variables
and $\langle n_d\rangle$ is a mean number of reservoir quanta (phonons),
related to the temperature $T$ as
$\langle n_d\rangle=\langle c^{\dag}(0) c(0)\rangle=
\left[ 1-\exp\left(\frac{\omega}{k T}\right)
\right]^{-1}$, where $k$ is the Boltzmann's constant.
From the Heisenberg-Langevin equations for $a$, $a^2$ and Hermitian
conjugated equations, we get  using Eqs. (\ref{5'}) and (\ref{6})
\begin{eqnarray}
\label{7}
i \frac{d}{d\tau}\langle\alpha\rangle & =&
\langle V\rangle-i\frac{\Gamma}{2}\langle\alpha\rangle, \nonumber\\
i \frac{d}{d\tau}\langle\left(\delta\alpha\right)^2\rangle &= &
2\langle V\delta\alpha\rangle+
\langle W\rangle
-i\Gamma\langle\left(\delta\alpha\right)^2\rangle, \\
i \frac{d}{d\tau}\langle\delta\alpha^*\delta\alpha\rangle & = &
-\langle V^*\delta\alpha\rangle+
\langle\delta\alpha^* V\rangle
-i\Gamma\langle\delta\alpha^*\delta\alpha\rangle
+i\Gamma\frac{\langle n_d\rangle}{N}, \nonumber
\end{eqnarray}
where
\begin{equation}
\label{W}
W(\alpha, \alpha^*)=\frac{1}{N} \frac{\partial V}{\partial a^{\dag}},\quad
z\equiv\langle a\rangle,\quad
\langle\left(\delta\alpha\right)^2\rangle=\langle a^2\rangle-z^2,\quad
\langle\delta\alpha^*\delta\alpha\rangle =
\langle a^{\dag} a\rangle-|z|^2,
\end{equation}
and we have introduced the scaled variables $\tau=g_l t$,
$\Gamma=\gamma/g_l$, $\bar{\Delta}=\Delta/g_l$.
Averaging in Eqs. (\ref{7}) is performed over both the reservoir variables
and the coherent state
$|\alpha\rangle=\exp(N\alpha a^{\dag} -N\alpha^* a) |0\rangle$
corresponding to the mean photon number $\simeq N$. Such kind of
minimum-uncertainty states are
most suitable for a consideration of the semiclassical limit $N\gg 1$
\cite{yaffe}. In derivation of Eqs.
(\ref{7}) we neglect an insufficient additional detuning introduced to
$\Delta$ by the interaction with the reservoir \cite{1,mandel_wolf}.
\par
The set of equations (\ref{7}) is not closed and actually is equivalent
to the infinite hierarchic dynamical system for the
cumulants of different order. To truncate it up to the
cumulants of the second
order, we make the substitution $a\rightarrow z +\delta\alpha$,
where at least initially the mean $z\simeq1$ and
the quantum correction $|\delta\alpha(0)|\simeq N^{-1/2}\ll 1$.
Using the Taylor expansion of the functions
$V(\alpha,\alpha^*)$ and $W(\alpha,\alpha^*)$ around the
mean values  $z$ and $z^*$,
\begin{equation}
\label{expansion}
V=V_z + \left(\frac{\partial V}{\partial\alpha}\right)_z
\delta\alpha +
\left(\frac{\partial V}{\partial\alpha^*}\right)_z
\delta\alpha^* +\cdots, \quad
 W=W_z + \left(\frac{\partial W}{\partial\alpha}\right)_z
\delta\alpha +
\left(\frac{\partial W}{\partial\alpha^*}\right)_z
\delta\alpha^* +\cdots,
\end{equation}
and after some algebra analogous to that used in \cite{9,9'},
we get  from (\ref{7}) in the first
order of $1/N$ the following self-consistent system
of equations for the mean value and the second-order cumulants
\begin{mathletters}
\label{details}
\begin{equation}
\label{details_a}
i \frac{d}{d\tau} z  = -i\frac{\Gamma}{2} z+\langle V\rangle_z +
\frac{1}{N} Q\left( z, z^*,\langle\left(\Delta\alpha\right)^2\rangle,
\langle\left(\Delta\alpha^*\right)^2\rangle,
\langle|\Delta\alpha|^2\right),
\end{equation}
\begin{equation}
\label{details_b}
i \frac{d}{d\tau}\langle\left(\Delta\alpha\right)^2\rangle =
2\left(\frac{\partial V}{\partial\alpha}\right)_z
\langle\left(\Delta\alpha\right)^2\rangle +
2\left(\frac{\partial V}{\partial\alpha^*}\right)_z
\langle|\Delta\alpha|^2\rangle + \langle w \rangle_z -
i\Gamma\langle\left(\Delta\alpha\right)^2\rangle,
\end{equation}
\begin{equation}
\label{details_c}
i \frac{d}{d\tau}\langle|\Delta\alpha|^2\rangle =
-\left(\frac{\partial V^*}{\partial\alpha}\right)_z
\langle\left(\Delta\alpha\right)^2\rangle +
\left(\frac{\partial V}{\partial\alpha^*}\right)_z
\langle\left(\Delta\alpha^*\right)^2\rangle-
i\Gamma\langle|\Delta\alpha|^2\rangle+
i\Gamma\langle n_d\rangle,
\end{equation}
\end{mathletters}
where we have introduced the scaled second-order cumulants as
$\langle\left(\Delta\alpha\right)^2\rangle=
N\langle\left(\delta\alpha\right)^2\rangle$ and
$\langle|\Delta\alpha|^2\rangle=
N\langle\delta\alpha^*\delta\alpha\rangle$, as well as the
function $w(z,z^*)=N W(z,z^*)$ (for definition of $W$ see Eq.
(\ref{W})). Now both the first-order cumulant $z$ and the
second-order cumulants
$\langle\left(\Delta\alpha\right)^2\rangle$,
$\langle|\Delta\alpha|^2\rangle$ are of the order of unity and
small parameter $1/N$ arises only as a prefactor for the quantum
correction $Q$ to the classical motion equation in (\ref{details_a}).
The expression for the quantum correction has the form of second
order differential $Q=\frac{1}{2} d^2 V\mid_z$.
In this formula and in Eqs. (\ref{details}), the subscript $z$ means that
the values of $V$ and its derivatives are calculated at the mean value $z$.
\par
The equations (\ref{details_b}), (\ref{details_c}) are nonlinear
due to the presence of the nonlinear term $\langle w \rangle_z$
in Eq. (\ref{details_b}). However, introducing new variables for
the second-order cumulants
\begin{equation}
\label{10}
B=\langle|\Delta\alpha|^2\rangle+\frac{1}{2},\quad
C=\langle\left(\Delta\alpha\right)^2\rangle,
\end{equation}
the term $\langle w \rangle_z$ could be removed and the
self-consistent equations (\ref{details}) may be rewritten as
\begin{mathletters}
\label{9}
\begin{equation}
i\dot{z}=-i\frac{\Gamma}{2} z+
\langle V\rangle_z + \frac{1}{N} Q(z, z^*, C, C^*, B),
\label{9a}
\end{equation}
\begin{equation}
i\dot{C}=2\left(\frac{\partial V}{\partial\alpha}\right)_z C +
2\left(\frac{\partial V}{\partial\alpha^*}\right)_z B
-i \Gamma C,
\label{9b}
\end{equation}
\begin{equation}
i\dot{B}=-\left(\frac{\partial V^*}{\partial\alpha}\right)_z C +
\left(\frac{\partial V}{\partial\alpha^*}\right)_z C^*
-i\Gamma\left( B-B^{(0)}\right),
\label{9c}
\end{equation}
\end{mathletters}
\begin{equation}
\label{10'}
Q(z, z^*, C, C^*, B)=
\frac{1}{2} \left(\frac{\partial^2 V}{\partial\alpha^2}\right)_z C +
\frac{1}{2} \left(\frac{\partial^2 V}{\partial\alpha^{*2}}\right)_z C^* +
\left(\frac{\partial^2 V}{\partial\alpha^{*}\partial\alpha}\right)_z
\left(B-\frac{1}{2}\right).
\end{equation}
If initially the oscillator is in the coherent
state, then the initial conditions for system
(\ref{9}) are
\begin{equation}
\label{initial_cond}
B(0)=1/2,\quad C(0)=0,
\end{equation}
and some arbitrary $z(0)\equiv z_0$ which is of the order of unity.
Involved into equation (\ref{9c}) an equilibrium value of cumulant
$B$ is determined by the mean number of reservoir's quanta and its
zero-point energy as
\begin{equation}
\label{B_0}
B^{(0)}=\langle n_d\rangle+1/2.
\end{equation}
Note that the zero-point energy of a reservoir appears in the
equations of motion
for the cumulants though it was not presented in the Heisenberg equations
of motion and even may be dropped from the Hamiltonian
redefining a zero of the energy. Such ``reappearence''
of a zero-point field energy is rather often in other problems of quantum
theory where a vacuum is responsible for the physical effects
\cite{milonni_book}.
\par
It is sufficient that now our basic equations for the
second-order cumulants (\ref{9b}), (\ref{9c}) are linear and this
fact allows us to find analytically the solution of whole set
(\ref{9}) using the quantum-classical correspondence in the limit
$N\rightarrow\infty$. The substitution (\ref{10}) describes, of
course, nothing but the transformation from a normal to a
symmetric ordering. The finding that equations of motion for the
lower-order cumulants in the semiclassical limit are solvable
just within the symmetric ordering is one of the illustration of
the fact that symmetric ordering is most suitable for the
consideration of the quantum-classical correspondence \cite{fabre}.
\par
Before we will find a solution of equations (\ref{9}),
let us make several comments and make a comparison with other approaches.\\
(i) It should be noticed that we derive the equations (\ref{7}), (\ref{9})
within the Heisenberg-Langevin approach, but alternatively, the same equations
could be obtained \cite{kna_98} starting from the corresponding generalized
Fokker-Planck equation \cite{1,mandel_wolf}.\\
(ii) In the case of no damping
$\Gamma=0$, our equations for the mean values and the second order cumulants
(\ref{9}) are reduced to the corresponding equations of ref. \cite{9,9'}.
On other hand, for the Hamiltonians which can be presented as sum of
a kinetic and a potential energy, our approach gives the same motion equations
as the cumulant expansion method introduced in \cite{sundaram}.
Moreover, Sundaram and Milonni showed \cite{sundaram} that the first-order
cumulant approximation to the Heisenberg equations of motion gives
the expectation values identical to those obtained by the methods
of Gaussian wavepacket
dynamics \cite{heller1,heller2}, semiquantum \cite{pattanayak,ashkenazy},
and based on the time-dependent
variational principle \cite{zhang}. In this respect, our present work could be
considered as some variant for the generalization of the semiquantum methods
to the case of a dissipative dynamics.\\
(iii)
Another approach to the description of the dynamics of quantum
fluctuations based on almost Gaussian wavepackets is the
so-called generalized Gaussian approximation (GGA) \cite{perina1,1,schack}.
This approximation is valid for both the Hamiltonian and the
dissipative systems and consists in an assumption
that the Fourier transform of quantum distribution function,
{\it i.e.} quantum
characteristic function, is Gaussian for any moment of time \cite{schack}.
Within GGA only the first- and the second-order cumulants are non zero,
and therefore it presents the
higher-order cumulants in terms of only the first and
the second-order cumulants and to
truncate an infinite dynamic system for the cumulants.
We compare our system (\ref{9}) with the corresponding dynamic
equations for the mean values and the cumulants
obtained within GGA for a several popular dissipative
models of quantum optics: a second harmonic generation \cite{szlachetka1},
a nondegenerate optical three-wave mixing\footnote{
The equations of motion within GGA for the model of nondegenerate three-wave
mixing could be easily obtained in an analogy with the case of second
harmonic generation \cite{szlachetka1}. The study of lossless three-wave
mixing within GGA has been presented in \cite{perina2}.
}
and
a forced nonlinear oscillator with cubic nonlinearity \cite{szlachetka2}.
For the forced nonlinear oscillator, we found that our
self-consistent set of
equations (\ref{9}) coincides with the corresponding basic equations
of \cite{szlachetka2} up to the terms of the order of $1/N^2$, and for the
problems of a nondegenerate and a degenerate optical 3-wave mixing, our
approach gives equations which are identical to the corresponding
motion equations obtained within GGA.
\par
There are yet the principal differences between our approach and
GGA: First, our approach works good in the semiclassical limit $N\gg 1$,
in contrast, GGA is suitable also for the quantum limit $N\simeq 1$. Second,
in the general case, the equations of motion obtained within GGA are
nonlinear and cannot be solved
analytically. As we shall see in the next subsection, the solution of our
system (\ref{9}) could be obtained analytically and has simple physical
meaning.
\subsection{The quantum-classical correspondence and the
dynamics of cumulants}
\label{subsec:Dynamics_of_cumulants}
We need to find the time dependencies of $z$, $C$, and $B$ from
the equations (\ref{9}).
First, suppose that we know the solution $z(\tau)$ of the
motion equation (\ref{9a}) for the mean value. Then,
the equations of motion for the second-order cumulants (\ref{9b})
and (\ref{9c})
form the linear inhomogeneous equation for the vector variable
${\bf X}(\tau)=\left[C(\tau), C^*(\tau), B(\tau)\right]$ as
\begin{equation}
\label{matrix}
i\dot{\bf X}=-i(\Gamma/2){\bf X}+\hat{A}{\bf X}+\epsilon{\bf X}_0,
\end{equation}
where ${\bf X}_0=(0, 0, i\Gamma  B^{(0)})$ and $B^{(0)}$ is given by
Eq. (\ref{B_0}). The matrix $\hat{A}$ is
formed by the partial derivatives of $V$ calculated at the mean value $z$,
its form can be easily obtained from Eqs. (\ref{9b}) and (\ref{9c})
(to save space we do not present an explicit form of $\hat{A}$ here).
We also introduced a parameter
$\epsilon$ for the discussion of solutions of Eq. (\ref{matrix})
(actual value $\epsilon\equiv 1$ in (\ref{matrix})).
The solution ${\bf X}(\tau)$ of linear Eq. (\ref{matrix}) consists of
two parts
\begin{equation}
\label{X_full}
{\bf X}(\tau)=\overline{\bf X}(\tau)+\tilde{\bf X}(\tau),
\end{equation}
where $\overline{\bf X}(\tau)$ is a
general solution of the homogeneous equation
(i.e., Eq. (\ref{matrix}) with $\epsilon=0)$ and
$\tilde{\bf X}(\tau)$ is a particular
solution of the inhomogeneous equation. It easy to see that
\begin{equation}
\label{X_tilde}
\tilde{\bf X}(\tau)=\left( 0, \Gamma B^{(0)}\tau\right).
\end{equation}
To find $\overline{\bf X}(\tau)$ we need to solve the self-consistent
set of equations (\ref{matrix}) for $\epsilon=0$ together with  (\ref{9a})
and (\ref{10'}).
We seek for a solution by the perturbation theory using the smallness
of the parameter $1/N$. Substituting the expression
\begin{equation}
\label{16}
z(\tau)=z_{cl}(\tau)+\frac{1}{N} z^{(1)}(\tau),\quad
\frac{1}{N} |z^{(1)}(\tau)|\ll |z_{cl}(\tau)|
\end{equation}
into the motion equation (\ref{9a}) for the mean value, we get in
the zero-order of $1/N$ the classical motion equation
\begin{equation}
\label{17}
i\dot{z}_{cl}=-i\frac{\Gamma}{2} z_{cl}+ V(z_{cl},z^*_{cl}).
\end{equation}
Now, it could be
shown that solution $\overline{\bf X}(\tau)$ of equation of motion
for the cumulants (\ref{matrix}) with $\epsilon=0$
can be obtained directly from the solution of the classical equation
(\ref{17}) by the linearization
near $z_{cl}$ (the substitution $z_{cl}\rightarrow z_{cl}+\delta z$,
$|\delta z|\ll |z|$), if one writes the dynamical equations for the variables
$(\delta z)^2$ and $|\delta z|^2$ (for the details of derivation, see
Appendix \ref{appendix1}).
Thus, the solution for $\overline{\bf X}(\tau)$ is
\begin{equation}
\label{X_overline}
\overline{\bf X}(\tau)=\left[ ({\rm d}z_{cl})^2,({\rm d}z^*_{cl})^2,
|{\rm d}z_{cl}|^2\right],
\end{equation}
where ${\rm d}z_{cl}(\tau)$ is the differential of the classical variable
$z_{cl}(\tau)$ governed
by Eq. (\ref{17}), and an initial conditions for (\ref{X_overline}) are
(\ref{initial_cond}), {\it i.e.} one should take
\begin{equation}
\label{initial_cond1}
({\rm d}z_{cl})^2(0)=0, \quad |{\rm d}z_{cl}|^2(0)=1/2.
\end{equation}
Now we can demonstrate how to find the dynamics of cumulants $C$ and $B$, if
the classical dynamics $z_{cl}(\tau)$ is known. Combining formulas
(\ref{X_full}), (\ref{X_tilde}), (\ref{X_overline}) and (\ref{B_0}),
we get
\begin{equation}
\label{B_and_C_general}
C=({\rm d}z_{cl})^2, \quad B=|{\rm d}z_{cl}|^2+
(\langle n_d\rangle+1/2)\Gamma\tau.
\end{equation}
Turn to the determination of the influence of the quantum correction
(\ref{10'}) on the dynamics
of the mean value $z$. Substituting the expansion (\ref{16}) into
equation (\ref{9a}), we have
in the first order of $1/N$ the motion equation for $z^{(1)}$ as
\begin{equation}
\label{z1_equation}
i\dot{z}^{(1)}=-i\frac{\Gamma}{2} z^{(1)}+ Q(\tau),\quad
Q(\tau)\equiv Q[z_{cl}(\tau), z^*_{cl}(\tau), C(\tau), C^*(\tau),
B(\tau)],
\end{equation}
where the cumulants $C(\tau)$, $C^*(\tau)$, and $B(\tau)$
are determined by the formulas (\ref{B_and_C_general}) and
$z_{cl}(\tau)$ is a solution of the classical equation (\ref{17}).
A formal solution of Eq (\ref{z1_equation}) is
\begin{equation}
\label{z1_solution}
z^{(1)}(\tau)=z^{(1)}(0) \exp\left(-\frac{\Gamma}{2}\tau\right)
-i\int_0^\tau d\tau' Q(\tau').
\end{equation}
For the initial conditions (\ref{initial_cond}), the initial value of
quantum correction (\ref{10'}) is zero. Therefore, $z(0)=z_{cl}(0)=z_0$
and $z^{(1)}(0)=0$. The condition of smallness of
$z^{(1)}(\tau)$ in comparison with $z_{cl}(\tau)$
[Eq. (\ref{16})] takes the form
\begin{equation}
\label{25}
R(\tau)\equiv
\left|\frac{z(\tau)-z_{cl}(\tau)}{z_{cl}(\tau)}\right|=
\frac{1}{N} \left| \frac{z^{(1)}(\tau)}{z_{cl}(\tau)}\right|=
\frac{1}{N} \left| \frac{\int_0^\tau d\tau' Q(\tau')}
{z_{cl}(\tau)}\right|\ll 1.
\end{equation}
We shall see when the condition (\ref{25}) is valid for the model
of anharmonic oscillator (\ref{1}) in the next section.
\par
We turn to the determination of the dynamics of the cumulants and
the mean value in the case of anharmonic oscillator with the Hamiltonian
(\ref{1}). Here, the expression for $V(z,z^*)$ takes the form
\begin{equation}
\label{26}
V(z,z^*)=\overline{\Delta} z +|z|^{2l}z.
\end{equation}
The exact solution of the classical motion equation (\ref{17}) with $V$
of the form (\ref{26}) is
\begin{equation}
\label{z_cl}
z(\tau)=z_0\exp\left[(-i\overline{\Delta} -\Gamma/2)\tau\right]
\exp\left[ -i |z_0|^{2 l}\mu_l(\tau)\right],\quad
\mu_l(\tau)=\left[ 1-\exp( -\Gamma l \tau ) \right]/\Gamma l.
\end{equation}
Using the expression (\ref{z_cl}), we have from Eqs. (\ref{B_and_C_general})
the following time dependencies for the cumulants
\begin{eqnarray}
\label{B_and_C}
C(\tau) &=&
-\mu_l(\tau) l z_0^2 |z_0|^{2(l-1)}\left( \mu_l(\tau) l |z_0|^{2 l}+i\right)
\exp\left[ (-\Gamma -i 2\overline{\Delta}) \tau -i 2 |z_0|^{2 l}
\mu_l(\tau) \right],
\nonumber\\
B(\tau) &=& \exp(-\Gamma\tau)\left[ 1/2+l^2 |z_0|^{4l}\mu_l^2(\tau) \right] +
\left( \langle n_d\rangle + 1/2\right) \Gamma\tau,
\end{eqnarray}
where we took into an account the initial conditions (\ref{initial_cond1}).
We shall use these time dependencies for the cumulants in the consideration
of the squeezing and the Fano factor in the next section.
\section{The squeezing and the Fano factor}
\label{sec:Squeez_and_Fano}
\subsection{The dynamics of squeezing}
\label{subsec:squeez}
Define the general field quadrature as $X_\theta=a\exp(-i\theta)+
a^{\dag}\exp(i\theta)$,
where $\theta$ is a local oscillator phase. A state is said to be squeezed
if there exists some $\theta$ such that
the variance of $X_\theta$ is smaller than the variance for a coherent
state or the vacuum
\cite{1,2}. Minimizing the variance of $X_\theta$ over $\theta$, we get
the condition for so-called principal squeezing \cite{1,2,4}
\begin{equation}
\label{sq_def}
S\equiv 1+2 N (\langle|\delta\alpha|^2\rangle-
|\langle(\delta\alpha)^2\rangle|)=2 (B-|C|) < 1.
\end{equation}
The determination of the principal squeezing $S$ is very useful because it
gives the maximal squeezing measurable by the homodyne
detection \cite{1,2}.
Substituting the expressions (\ref{B_and_C}) for the cumulants $B$ and $C$
into the definition (\ref{sq_def}), we find
\begin{equation}
\label{30}
S(\tau)=\exp(-\Gamma\tau) \left[1+\phi_l(x_0,\tau)\right]
+(\langle n_d\rangle +1/2) 2\Gamma\tau,
\end{equation}
\begin{equation}
\label{31}
\phi_l(x_0,\tau)=2a\left[ a-(1+a^2)^{1/2}\right],\quad
a(\tau)=l x_0^{2l}\mu_l(\tau),
\end{equation}
where $\mu_l(\tau)$ is defined in (\ref{z_cl}) and
we assumed for the sake of simplicity that the initial condition $z_0$
is real $x_0={\rm Re} z_0$. First, consider some interesting particular
cases.
\subsubsection{The weak dissipation}
In the case of weak dissipation $\Gamma\tau\ll 1$, we have
$\exp(-\Gamma\tau)\approx 1 -\Gamma\tau$ and  $\mu_l(\tau)\approx\tau$.
Thus, in this limit, from Eqs. (\ref{30}) and (\ref{31}), we get
\begin{equation}
\label{34}
S(\tau)=1+(1-\Gamma\tau)\phi_l(x_0,\tau)+2\langle n_d\rangle\Gamma\tau,
\quad a(\tau)\approx l x_0^{2l}\tau\equiv l {\cal P}^{(2l+1)} t.
\end{equation}
In the case of no loss ($\Gamma=0$), the formula (\ref{34})
shows that the rate of squeezing is determined by the factor
$2 l x_0^{2 l}\lambda_l N^l\equiv 2 l {\cal P}^{(2 l+1)}$. Because of
$\lambda_l$ is proportional to the $(2 l+1)$-order nonlinear
susceptibility, the factor ${\cal P}^{(2 l+1)}$ has a physical meaning
of the nonlinear polarization. Therefore, the stronger is nonlinear
polarization induced by a light in the medium, the more effective
squeezing of light is possible.
For a finite dissipation $\Gamma\not=0$,
the squeezing is determined by an interplay between dissipation,
the polarization of the nonlinear medium modelled by the anharmonic
oscillator and the thermal fluctuations of a reservoir. Both dissipation
and a finite temperature of reservoir are the destructive factors for
the squeezing. It is interesting to note that the vacuum term does not appear
in Eq. (\ref{34}), what is characteristic of the weak dissipation limit.
\subsubsection{ The short-time approximation $\tau\ll 1$}
The short-time approximation $\tau\ll 1$, as well as the limit of
large photon number $N\gg 1$,
are quite realistic for a nonlinear medium modelled by the
anharmonic oscillators
(for numerical estimates, see \cite{1}, chap. 10, and \cite{4}).
In the limits of $\tau\ll 1$ and $\Gamma\tau\ll 1$,
we have from Eq. (\ref{31}): $\phi_l(x_0,\tau)\approx -2l x_0^{2l}\tau=
-2l {\cal P}^{(2 l+1)} t$. Thus, substituting this expression
into (\ref{34}), we get a very simple dependence of $S$ on time as
\begin{equation}
\label{35}
S(t)=1-\left[ l x_0^{2l}-\Gamma\langle n_d\rangle\right]
2\tau+O(\tau^2).
\end{equation}
As follows from Eq. (\ref{35}), there exists the critical number of phonons
$n_d^{(cr)}=(l/\gamma) {\cal P}^{(2 l+1)}$
such that at $n_d\ge n_d^{(cr)}$ any degree of squeezing is impossible.
\subsubsection{The lossless case $\Gamma=0$}
In the case of no loss, the time dependence of $S$ is
\begin{equation}
\label{32}
S(\tau)=1+\phi_l(x_0,\tau),
\end{equation}
where $\phi_l$ is defined in (\ref{31}) and we should take
into account
that now $\mu_l(\tau)=\tau$ and thus $a(\tau)=l x_0^{2l}\tau=
l {\cal P}^{(2l+1)} t$.
For the case of Kerr nonlinearity ($l=1$), the formula (\ref{32})
coincides with the corresponding formula from \cite{4} obtained in the
limit $N\gg 1$ from the exact solution for the Hamiltonian case.
\par
Now we consider the time dependence of $S$ [Eq. (\ref{32})] in the limit
of large time $\tau\gg 1$, which corresponds to $a\gg 1$.
Rewriting $\phi_l$ in the form
$\phi_l(x_0,\tau)=2 a^2 \left[ 1- (1+a^{-2})^{1/2}\right]$ and
expanding $(1+a^{-2})^{1/2}$ up to the term of order of $a^{-3}$,
we get $\phi_l\approx -1+(6 a)^{-1}$ for $a\gg 1$. Therefore
\begin{equation}
\label{33}
S(\tau)\approx \left( l x_{0}^{2l}\tau\right)^{-1}
=\left( l {\cal P}^{(2l+1)} t \right)^{-1}
\end{equation}
for $\tau\gg 1$. This is in good agreement with the statement that
the principal squeezing $S$ is a power-wise decreasing function of time
for the general class of integrable systems in the semiclassical limit
\cite{9'}.
\par
Return to the case of an arbitrary dissipation.
The dependence of $S$ on $\tau$ [Eqs. (\ref{30}), (\ref{31})] is shown in
Fig.~1, where the solid curve corresponds to the lossless case ($\Gamma=0$),
the dashed curve corresponds to the finite damping constant
$\Gamma=5\times 10^{-2}$ with a reservoir
of zero temperature $\langle n_d\rangle=0$, and, finally, the dotted curve
corresponds to the reservoir with finite temperature
($\langle n_d\rangle=1$).
As is evident from this figure, the squeezing is stronger for
higher nonlinearity (compare Fig.~1a and Fig.~1b),
a dissipation slows down the rate of squeezing
(compare the solid and the dashed curves in Fig.~1).
Moreover, a finite temperature of a reservoir fastly destroys the squeezing.
\subsection{The Fano factor}
\label{subsec:Fano}
Another important characteristic of the nonclassical properties of
a light is the Fano factor
\begin{equation}
\label{Fano_1}
F=\frac{\langle n^2\rangle-\langle n\rangle^2}{\langle n\rangle},
\end{equation}
which determines the deviation of a probability distribution from
the Poissonian distribution with $F=1$ \cite{1,2}.
In Eq. (\ref{Fano_1}) the mean foton number $\langle n\rangle$ and
the mean of square of foton number $\langle n^2\rangle$  are
\begin{equation}
\label{Fano_2}
\langle n\rangle\equiv \langle b^{\dag} b\rangle=
N\langle a^{\dag} a\rangle,\quad
\langle n^2\rangle=N^2\langle a^{\dag} a a^{\dag} a\rangle=
N^2\langle a^{\dag 2} a^2\rangle+\langle n\rangle.
\end{equation}
Substituting expression $a\rightarrow z +\delta\alpha$
$(|\delta\alpha|\ll |z|\simeq 1)$
into formulas (\ref{Fano_2}),
we have after Taylor expansions in the first order of $1/N$
\begin{equation}
\label{Fano_3}
\langle a^{\dag} a\rangle=|z|^2 + N^{-1} (B-1/2),\quad
\langle a^{\dag 2} a^2\rangle =|z|^4 + N^{-1} (z^* C + {\rm c.c.})
+N^{-1} 4 |z|^2 (B-1/2),
\end{equation}
and, therefore, the dependence of the Fano factor on the cumulants and
the mean values in the first order of $1/N$ is\footnote{
Note that this expression coincides with the corresponding result
obtained within GGA (see formulas (3.153) and (10.42) in \cite{1})
up to the terms of order of $1/N^2$.
}
\begin{equation}
\label{Fano_4}
F=2 B+\left(\frac{z^*}{z} C+ {\rm c.c.}\right).
\end{equation}
Substituting expressions (\ref{B_and_C}) for $B$ and $C$  into
Eq. (\ref{Fano_4}), we find the following time dependence for the
Fano factor
\begin{equation}
\label{Fano_5}
F(\tau)=\exp(-\Gamma \tau)+\left( \langle n_d\rangle+1/2
\right) 2\Gamma\tau.
\end{equation}
As it is evident from Eq. (\ref{Fano_5}), the statistics is the
super-Poissonian $F(\tau)>1$ at any time.
The most simple form the time dependence for the Fano factor takes
in the case of weak dissipation
\begin{equation}
\label{Fano_6}
F(\tau)=1+2\langle n_d\rangle\Gamma\tau, \quad \Gamma\tau\ll 1.
\end{equation}
Thus, the statistics is super-Poissonian for any $\Gamma\not=0$ and is
independent on the degree of nonlinearity $l$.
This is in good agreement with the previous result of \cite{2,6} for
the case of the dissipative Kerr oscillator ($l=1$), where from
the exact solution the impossibility of sub-Poissonian statistics and
the antibunching has been predicted.
\section{The conditions of validity of the $1/N$-expansion
          and the quantum-classical correspondence}
\label{sec:validity}
The procedure of the $1/N$-expansion may be considered
self-consistently, if the influence of the quantum correction to
classical motion on the dynamics of the mean values is small, {\it i. e.}
the condition (\ref{25}) is satisfied.
From Eqs. (\ref{10'}) and (\ref{26}), we have the following
expression for the quantum correction $Q$ in the case of an arbitrary
nonlinearity $l$
\begin{equation}
\label{Q_osc}
Q(z,z^*)=\frac{1}{2}l(l+1) z^{*l} z^{l-1}  C+
\frac{1}{2}l(l-1) z^{*(l-2)} z^{l+1}  C^* +
l(l+1) z^{*(l-1)} z^l (B-1/2).
\end{equation}
We start the consideration of the condition of validity (\ref{25})
with some particular cases.
\subsubsection{The short-time approximation}
First, we consider the short-time
approximation. In this case, the integral over $Q(\tau)$ in (\ref{25})
could be replaced by the product and we have
\begin{equation}
\label{short_appl}
\left|\frac{z(\tau)-z_{cl}(\tau)}{z_{cl}(\tau)}\right|=
\frac{\tau}{N} \left| \frac{Q(\tau)}{z_{cl}(\tau)}\right|\ll 1.
\end{equation}
It is easy to see that for $\tau\ll 1$, $Q(\tau)\simeq Q(0)$
and thus $|(z-z_{cl})/z_{cl}|$ is of the order of $1/N$.
Therefore, the condition of validity
within the short-time approximation (\ref{short_appl}) is fulfilled for
any number of photons $N\gg 1$ large enough.
\subsubsection{The lossless case $\Gamma=0$}
In Appendix  \ref{appendix2} we show that the time interval of validity
of our approach for the lossless case $\Gamma=0$ is
\begin{equation}
\label{lossless_appl}
\tau\ll\tau^*_{ham}=N^{1/2}\frac{|\bar{\Delta}+|z_0|^{2l}|^{1/2}}
{l^{3/2}|z_0|^{3l-1}}.
\end{equation}
This result is in good agreement with our previous finding \cite{9'}
that for the Hamiltonian integrable models the time scale for the
validity of the $1/N$-expansion has a power-wise dependence in $N$
in the semiclassical limit $N\gg 1$. Now we turn to the general case
$\Gamma\not=0$.
\subsubsection{The case of an arbitrary dissipation}
The condition of validity (\ref{25})
cannot be obtained analytically for an arbitrary time and for
an arbitrary damping $\Gamma$, but combining the computations and
simple analytic estimates for some limiting case, we can
qualitatively understand the behavior of $R(\tau)$ [Eq.
(\ref{25})]. We start with the time dependence of the quantum
correction $Q(\tau)$ [Eqs. (\ref{z1_equation}), (\ref{Q_osc}),
(\ref{z_cl}), (\ref{B_and_C})]. This function is shown in Fig.~2
for the different degrees of nonlinearity (the solid line for
$l=1$ and the dashed line for $l=3$), as well as for a rather
week (Fig.~2a) and for a relatively strong (Fig.~2b) damping
constants. The quantum correction $Q(\tau)$ first increases, then
reaches some maximum and finally decreases. The maximum of
$Q$ increases with the increase of nonlinearity degree (compare
the solid and the dashed curves in Fig.~2) and shifts to a shorter
time with the increase of dissipation (compare Fig.~2a and
Fig.~2b).
\par
Such behavior is in a sharp contrast to the lossless
(Hamiltonian) nonlinear systems, where the quantum correction
always increases with time either power-wise for regular
dynamics \cite{berman-book} or exponentially for chaotic dynamics
\cite{berman-book}. Before $Q(\tau)$ gets to a maximum, during
some time interval $[0,\tau_1]$ it could be well approximated by
the time dependence for the lossless case
$Q(\tau<\tau_1)\simeq l^3\tau^2$ (see formula (\ref{B8})). If we
estimate the time scale $\tau_1$ from the behavior of the most
fastly decreasing function $\mu_l(\tau)$ [Eq. (\ref{z_cl})],
we have
\begin{equation}
\label{tau_1}
\tau_1=\left(\Gamma l\right)^{-1},\quad
Q(\tau_1)\simeq l^3\tau_1^2=l/\Gamma^2
\end{equation}
for the position and the value of maximum of $Q(\tau)$. As is easy to
see from Fig.~2, this naive estimate, however, well represents
all main features of the time behavior of $Q(\tau)$.
\par
Consider now the time dependence of the difference between the
classical and the quantum mean values
$z^{(1)}(\tau)=N|z(\tau)-z_{cl}(\tau)|$
caused by the existence of a quantum correction to the classical
motion. Computing the integral over $Q(\tau)$ [Eq.
(\ref{z1_solution})], we plot the time dependence $z^{(1)}(\tau)$
in Fig.~3 for weak (Fig.~3a) and for strong (Fig.~3b)
damping, as well as for the different degrees of nonlinearity.
This plot shows that the difference $z^{(1)}(\tau)$ first increases
and then saturates at some level determined by the degree of
nonlinearity and dissipation.
\par
We start our analysis with the case of weak dissipation
(Fig.~3a). For weak dissipation, the level of saturation of
$z^{(1)}(\tau)$ increases with consequent consideration of higher
nonlinearities. This result may be qualitatively understood as
follows. For weak dissipation, the time interval of almost
lossless behavior $\tau\lesssim\tau_1$ (\ref{tau_1}) is rather
long and therefore the asymptotic behavior of $z^{(1)}(\tau)$
sufficiently depends on its behavior during the time
$0\leq\tau\lesssim\tau_1$. For $\tau\lesssim\tau_1$, we can use
the approximation (\ref{B9}), i.e. $|z^{(1)}(\tau_1)|\simeq\tau_1^2
l^3=l/\Gamma^2$. This estimate shows that the level of saturation
for $z^{(1)}(\tau)$ should increase with the growth of $l$.
\par
Turn to the case of strong dissipation (Fig.~3b). For strong
dissipation, the level of saturation of $z^{(1)}(\tau)$ is
sufficiently lower in comparison with the case of weak
dissipation (compare Fig.~3a and Fig.~3b). Moreover, the level of
saturation for $l=1$ is slightly higher than for nonlinearities
with $l>1$ and the difference of the saturation levels for the different
nonlinearities with $l>1$ is practically invisible. To understand
such behavior we note that for strong enough dissipation the
time interval for almost lossless dynamics $\tau\lesssim\tau_1$
is short and the behavior of $Q(\tau)$ for another time interval
$[\tau_1, \infty]$ is more sufficient (Fig.~2b). For the
nonlinearity with $l=1$, $Q(\tau)$ has slowly decreasing tail,
while it decays rapidly for $l>1$ (Fig.~2b). As a result, the
area under the curve $Q(\tau)$ for $l=1$ is greater than the
areas under the curves corresponding to different $l$ with $l>1$.
Yet, because all $Q(\tau)$ are fastly decreasing
functions at $\tau>\tau_1$ for all $l>1$, the levels of saturation
of $z^{(1)}(\tau)$ corresponding to different $l$ are indistinguishable in
the scale of Fig.~3.
\par
Till now we studied the time dependence of the difference
between quantum mean value and classical solution
$|z(\tau)-z_{cl}(\tau)|$. However, the criterion of validity of
the $1/N$-expansion in the form (\ref{25}) includes also the
$z_{cl}(\tau)$ as $R(\tau)=N^{-1}|z^{(1)}/z_{cl}|$. During the time
interval of order of $\tau_1$, when $z^{(1)}(\tau)$ is growing,
$z_{cl}$ is slowly decreasing and oscillating function. But when
$z^{(1)}(\tau)$ saturates, simultaneously the classical solution should be
considered as a rapidly decreasing function
$z_{cl}\simeq\exp(-\Gamma\tau/2)$. Therefore, for the plateaus in
Fig.~3, $R(\tau)\simeq N^{-1}\exp(\Gamma\tau/2)$ resulting in the
corresponding time scale for validity of our approach on a
plateau as
\begin{equation}
\label{log}
\tau<\tau^*=\Gamma^{-1}\ln N.
\end{equation}
Formally, this result looks like corresponding estimate for the
time scale for validity of the semiclassical description
(``breaking time'') in the
quantum chaotic systems $\tau^*=\lambda^{-1}\ln N$, where
$\lambda$ is the maximal Lyapunov exponent
\cite{berman-book,sundaram,8,9,9'}.
However, the physical reason for the appearance of a very short, logarithmic
breaking time is quite different in our case. When a classical
oscillator spiraled around an equilibrium state, the amplitude of
the oscillations becomes very small and the quantum description
starts to differ significantly from classical one due to the
uncertainty principle. That is the reason for the logarithmically
short time scale of the semiclassical description of strongly
damped oscillations.
\par
However, the behavior of an oscillator near an equilibrium point
is a physically not very interesting process. On the opposite, the
physically interesting case when the oscillations are not
overdamped, we can again apply the estimate (\ref{B9}) during
the time interval $\tau\simeq\tau_1=(\Gamma l)^{-1}$ and get the
criterion of validity in the form
\begin{equation}
\label{loss_appl}
R(\tau)=\frac{1}{N} \left|\frac{z^{(1)}(\tau_1)}{z_{cl}(\tau_1)}\right|
\simeq\frac{l}{N\Gamma^2}\ll 1.
\end{equation}
For large enough $N$, the criterion (\ref{loss_appl}) could be
always satisfied and our approach works well.
\par
We can summarize our findings concerning the quantum-classical
correspondence and the conditions for validity of the
$1/N$-expansion as follows. In a lossless case, the quantum
correction to the classical motion equation grows quadratically
with time. For finite dissipation, the correction $Q(\tau)$
first increases and then decreases with time. The difference
between the solution for the quantum expectation value and the
classical solution $|z(\tau)-z_{cl}(\tau)|$ first increases a
power-wise with time and then saturates at some level, which is
dependent on the degree of nonlinearity and on the dissipation
strength. The condition for validity has different
scaling dependence on the semiclassical parameter $N$ for two
different time intervals. During the time interval of growth of
$|z(\tau)-z_{cl}(\tau)|$, the criterion of validity has the form
(\ref{loss_appl}), i.e. $R\simeq 1/N\ll 1$, and it is certainly
satisfied for large $N$. During the time interval when
$|z(\tau)-z_{cl}(\tau)|$ saturates, our approach is valid only
for a short time $\tau^*\simeq\ln N$. We believe that described
above conditions for the semiclassical description of the quantum
system are common for all dissipative systems with simple
attractor.
\section{Conclusion}
\label{sec:Conclusion}
In summary, using the quantum-classical correspondence, we find
analytically the time dependencies of squeezing and the Fano
factor for the dissipative anharmonic oscillators of an
arbitrary degree of nonlinearity in the limit of large number of
fotons. It should be noticed that our basic equations are rather
general and could be considered as some generalization of the
Gaussian wavepacket dynamics methods to the dissipative systems.
Moreover, some effects mentioned in the present work are
also rather general. For instance, quantum diffusion around
classical trajectory in the dissipative system, which we
interpret as the influence of the reservoir's vacuum on the quantum
system, should be observable not only in the dissipative
nonlinear oscillators. In this respect, we would like to mention
the work of Savage published ten years ago \cite{savage}. The
author of \cite{savage} studied numerically, within the
Gaussian approximation, the
quantized version of the second harmonic generation problem for
the self-oscillating regime \cite{mcneil}. He had found
numerically a diffusive growth of the quantum variances near a
classical limit cycle. However, neither an analytic treatment of the
problem, nor a physical explanation of this ``diffusion'' were
presented in \cite{savage}. We plan to devoted our future
publication to the investigation of this problem together with more
accurate study of the validity conditions for the application of
the $1/N$-expansion to the dissipative quantum systems.
\section*{Acknowledgments}
Discussions with Evgeny Bulgakov, Antoine Heidmann, Vlasta
Pe\v{r}inov\'{a} are gratefully acknowledged. We also thank
Serge Reynaud for stimulating interest to the activity.
KNA thanks Department of Optics, Palack\'{y} University and
Joint Laboratory for Optics, Olomouc, as well as Department of Physics,
University of Illinois at Urbana-Champaign for hospitality during
the work on this project.
This work was partially supported by Czech Grant Agency,
grant No 202/96/0421, Czech Ministry of Education, grant No VS96028
and Russian Fund for Basic Research, grant No 96-02-16564.
\appendix
\section{}
\label{appendix1}
In this Appendix we show that the equations of motion for the cumulants
(\ref{matrix}) at $\epsilon=0$ could be obtained from the classical
motion equation (\ref{17}). This proof is valid for any dissipative system
with one degree of freedom.
\par
Linearization of classical equations (\ref{17}) near $z_{cl}$  by
means of the substitution $z_{cl}\rightarrow z_{cl}+\delta z$ gives
\begin{equation}
\label{A1}
i \frac{d}{d\tau}\delta z  =
-i\frac{\Gamma}{2} \delta z +
\frac{\partial V}{\partial z} \delta z +
\frac{\partial V}{\partial z^*}\delta z^*,\quad
i \frac{d}{d\tau}\delta z^*  =
-i\frac{\Gamma}{2} \delta z^*
-\frac{\partial V^*}{\partial z}\delta z -
\frac{\partial V^*}{\partial z^*}\delta z^*,
\end{equation}
where all derivatives are taken on the classical trajectory $z_{cl}(\tau)$.
Using the equality
$\left( \partial V/\partial z\right)=
\left( \partial V^* /\partial z^* \right)$,
we have from Eq. (\ref{A1}) for the quadratic variables $(\delta z)^2$ and
$|\delta z|^2$ the following equations of motion
\begin{eqnarray}
\label{A2}
i \frac{d}{d\tau}\left(\delta z\right)^2 &=&
-i\Gamma \left(\delta z\right)^2 +
2\frac{\partial V}{\partial z} \left(\delta z\right)^2 +
2\frac{\partial V}{\partial z^*} |\delta z|^2,\nonumber \\
i \frac{d}{d\tau} |\delta z|^2 &=&
-i\Gamma |\delta z|^2
-\frac{\partial V^*}{\partial z} \left(\delta z\right)^2 +
\frac{\partial V}{\partial z^*} \left(\delta z^*\right)^2.
\end{eqnarray}
It is easy to see that Eqs. (\ref{A2}) are equivalent to
Eqs. (\ref{matrix}) with $\epsilon=0$,
if one makes substitutions
$|\delta z|^2\rightarrow B$ and $(\delta z)^2\rightarrow C$.
\par
 It should be noticed that yet exists one difference
 between the linearization of the classical motion
equations and the equations for quantum cumulants (\ref{9b}), (\ref{9c}):
It is impossible
to get the initial conditions (\ref{initial_cond}) for $C$ and $B$
from only initial conditions for the linearized classical equations of
motion (see also discussion of this problem in \cite{9,9'}).
\section{}
\label{appendix2}
In this Appendix, we present the derivation of the time scale for the validity
of the $1/N$-method for lossless ($\Gamma=0$) oscillators.
Start with the case of Kerr nonlinearity $l=1$ and then generalize
obtained results for arbitrary $l$. From
Eqs. (\ref{z_cl}) and (\ref{B_and_C}) for $\Gamma=0$ and $l=1$, we get
\begin{equation}
\label{B1}
z_{cl}(\tau)=z_0\exp(-i\Omega\tau), \quad
\Omega\equiv\bar{\Delta}+|z_0|^2,
\end{equation}
\begin{equation}
\label{B2}
C(\tau)=-\tau z_0^2 \left[ |z_0|^2\tau+i\right]
\exp(-i 2\Omega\tau), \quad
B(\tau)=1/2+|z_0|^4 \tau^2.
\end{equation}
The expression (\ref{Q_osc}) for quantum correction in this case is
\begin{equation}
\label{B3}
Q(z, z^*, C, C^*, B)=z^* C+2 z B -z.
\end{equation}
Substituting (\ref{B1}) and (\ref{B2}) into (\ref{B3}), we have
for defined in Eq. (\ref{z1_equation}) function $Q(\tau)$:
\begin{equation}
\label{B4}
Q(\tau)=z_0 |z_0|^4 \tau^2 \exp(-i\Omega\tau)
-i z_0 |z_0|^2 \tau \exp(-i\Omega\tau),
\end{equation}
and involved in (\ref{25}) integral is
\begin{eqnarray}
\label{B5}
\int_0^\tau d\tau' Q(\tau') &=& z_0 |z_0|^4\Omega^{-3}\left[
2\Omega\tau\exp(-i\Omega\tau)+i (\Omega^2\tau^2-2)\exp(-i\Omega\tau)
+2 i \right] - \nonumber\\
 & & i z_0 |z_0|^2 \Omega^{-2}\left[
\exp(-i\Omega\tau)+i \Omega\tau\exp(-i\Omega\tau)-1\right].
\end{eqnarray}
The most rapidly increasing term in (\ref{B5}) for $\tau\gg 1$ is
$\Omega^{-1}z_0 |z_0|^4\tau^2\exp(-i\Omega\tau)$. Substituting this
expression into Eq. (\ref{25}), we have condition of validity in the form
\begin{equation}
\label{B6}
\frac{|z_0|^4}{|\bar{\Delta}+|z_0|^2|}\frac{\tau^2}{N}\ll 1,
\end{equation}
and therefore the time scale $\tau^*$ of validity is
\begin{equation}
\label{B7}
\tau\ll\tau^*=N^{1/2}\frac{|\bar{\Delta}+|z_0|^2|^{1/2}}{|z_0|^2}.
\end{equation}
Turn to the lossless case for an arbitrary $l$. Analogously to
the case $l=1$, it could be shown that the expression for quantum correction
(\ref{z1_equation}) in the limit of large time $\tau\gg 1$ is
\begin{equation}
\label{B8}
Q(\tau)\approx z_0 |z_0|^{2(3l-1)} l^3 \tau^2\exp[-i\Omega_l\tau], \quad
\Omega_l\equiv\bar{\Delta}+|z_0|^{2l},
\end{equation}
and the influence of quantum correction on mean value (\ref{z1_solution}) is
\begin{equation}
\label{B9}
z^{(1)}(\tau)=\int_0^\tau d\tau' Q(\tau')\approx
\frac{l^3 z_0 |z_0|^{2(3l-1)} \exp(-i\Omega_l\tau)}{\Omega_l} \tau^2,
\end{equation}
where we again presented only the term which is leading in time for $\tau\gg 1$.
Finally, substituting (\ref{B9}) to (\ref{25}) with account of (\ref{z_cl}),
we obtain (\ref{lossless_appl}).
%

%
\vspace{4cm}
\epsfxsize=17cm
\hspace{0.5cm}
\epsfbox{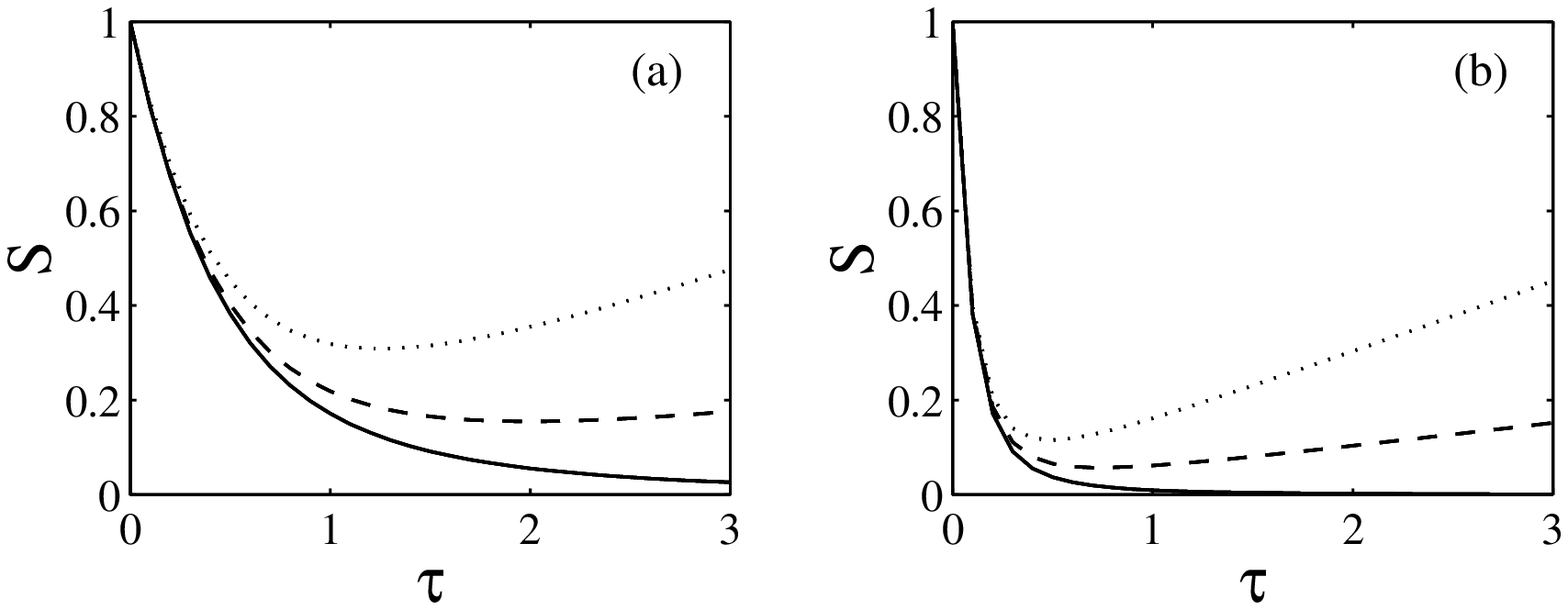}
\begin{figure}
\caption{The dependence of squeezing $S$ on scaled time $\tau$
for the Kerr nonlinearity with $l=1$ (a) and with $l=5$ (b).
Other parameters are following:
$\Gamma=0$, $\langle n_d\rangle=0$ (solid curve);
$\Gamma=5\times 10^{-2}$, $\langle n_d\rangle=0$ (dashed curve);
and $\Gamma=5\times 10^{-2}$, $\langle n_d\rangle=1$ (dotted curve).
The initial condition is $z_0=1$.}
\label{fig1}
\end{figure}
%
%
\vspace{3cm}
\epsfxsize=17cm
\hspace{0.5cm}
\epsfbox{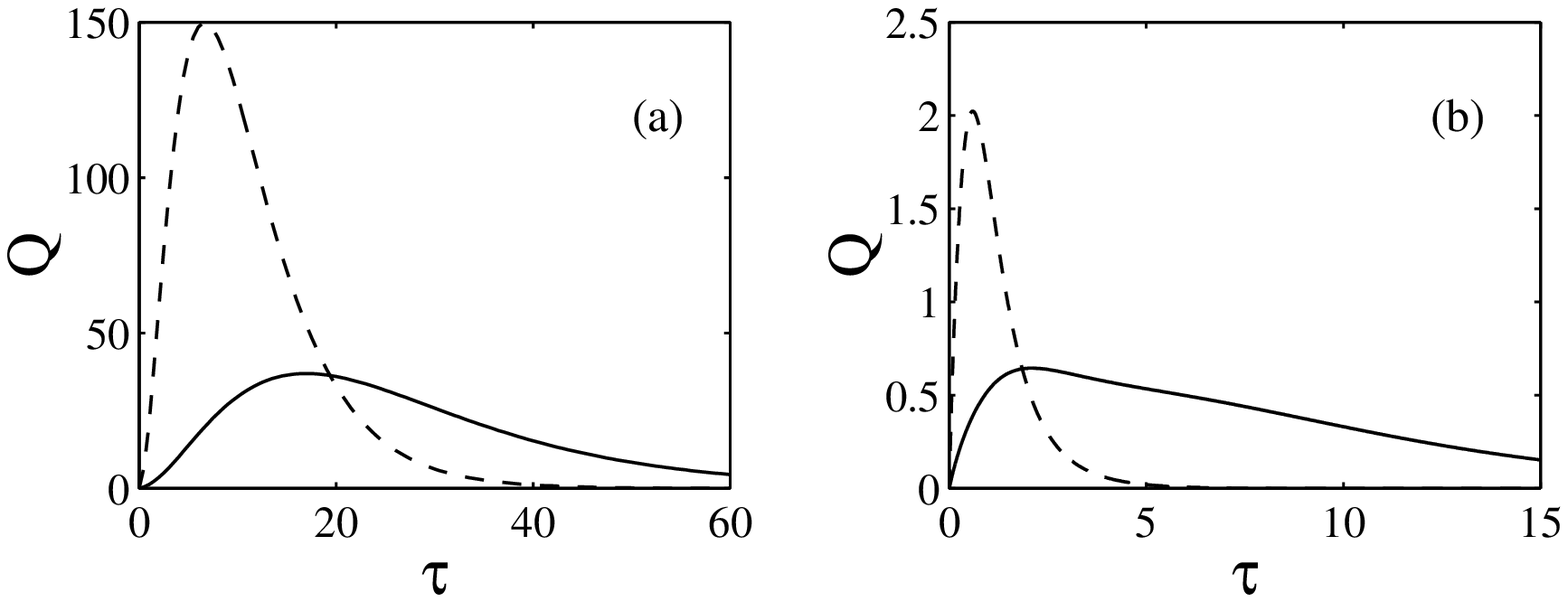}
\begin{figure}
\caption{The dependence of the quantum correction
$Q$ on scaled time $\tau$ for weak damping $\Gamma=0.05$ (a),
and for strong damping $\Gamma=0.5$ (b). Solid curve corresponds
to the Kerr nonlinearity with $l=1$ and dashed curve -- to nonlinearity
with $l=3$. Everywhere $\langle n_d\rangle=0$, $\Delta=0$, $z_0=1$.}
\label{fig2}
\end{figure}
%
%
\vspace{5cm}
\epsfxsize=18cm
\hspace{0.5cm}
\epsfbox{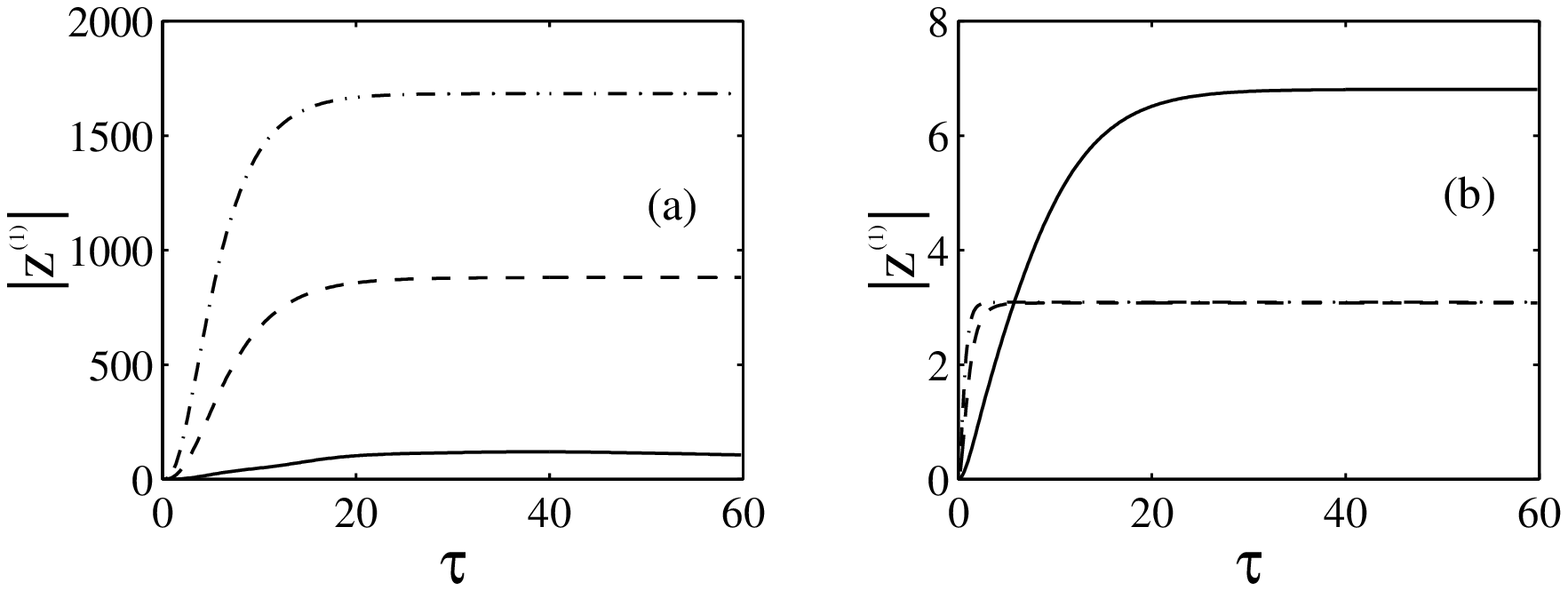}
\begin{figure}
\caption{The scaled difference between the quantum and the classical
expectation values $|z^{(1)}|=N |z-z_{cl}|$ plotted as a function of
time $\tau$ for weak damping $\Gamma=0.05$ (a),
and strong damping $\Gamma=0.5$ (b). The Kerr nonlinearity with $l=1$ is
described by the solid curve, dashed curve corresponds to the nonlinearity with
$l=3$, dashed and dotted curve corresponds to the nonlinearity with $l=5$.
Other parameters and initial condition are the same as in Fig. 2.}
\label{fig3}
\end{figure}
\end{document}